\documentclass[12pt,preprint]{aastex}

\def\lessim{\mathrel{\hbox{\rlap{\hbox{\lower4pt\hbox{$\sim$}}}\hbox{$<$}}}}
\def\grtsim{\mathrel{\hbox{\rlap{\hbox{\lower4pt\hbox{$\sim$}}}\hbox{$>$}}}}

\shorttitle{Nova in M31 Globular Cluster}
\shortauthors{Shafter et al.}

\begin{document}


\title{M31N-2007-06b: A Nova in the M31 Globular Cluster Bol~111}


\author{A. W. Shafter}
\affil{Department of Astronomy, San Diego State University,
    San Diego, CA 92182}
\email{aws@nova.sdsu.edu}

\and

\author{R. M. Quimby\altaffilmark{1}}
\affil{University of Texas, Austin TX, 78712}
\email{quimby@phobos.caltech.edu}


\altaffiltext{1}{Present Address: California Institute of Technology}


\begin{abstract}
We report spectroscopic observations of the nova M31N-2007-06b, which
was found to be spatially coincident
with the M31 globular cluster Bol 111. This nova is the
first out of more than 700 discovered in M31 over the past century
to be associated with one of the galaxy's globular clusters. A total of three
spectra of the nova were obtained 3, 6, and 36 days post discovery. The
data reveal broad (FWHM$\sim3000$~km~s$^{-1}$) Balmer, NII, and NIII
emission lines, and show that the nova belongs to
the He/N spectroscopic class. The He/N class of novae are relatively
rare, making up roughly 15\% of the novae with measured spectra in
M31, and roughly 20-25\% of the Galactic novae for which spectroscopic
data are available. The implications of a nova, particularly an He/N nova,
occurring in a globular cluster are discussed.

\end{abstract}




\section{Introduction}

Classical Novae are semi-detached binary stars where the Roche-lobe-filling
component (typically a cool, near-main-sequence star) transfers
mass to a white dwarf companion (Warner 1995). If the rate of accretion is
low enough to allow the accreted material to become sufficiently degenerate,
a thermonuclear runaway (TNR) eventually ensues, driving substantial
mass loss from the system, and leading to a nova eruption.
Luminosities as high as M$_V \simeq -9$ are often observed
at the peak of eruption, making novae
second only to gamma-ray bursts and supernovae in the
energetics of their outbursts.
Their high luminosities ($-6 \lessim$ M$_V \lessim -9$)
and frequencies of occurrence
($\sim50$~yr$^{-1}$ in a galaxy like M31),
make them potentially powerful
probes of the evolution of binary systems in different (extragalactic) stellar
populations.

The standard picture for the formation of a nova system is through the
common envelope evolution of an initially detached
progenitor binary leading to the formation
of a short-period, semi-detached system with the
red dwarf transferring mass to its white dwarf companion (Paczy\'nski 1976;
Meyer \& Meyer-Hofmeister 1979).
Population synthesis studies based on this formation mechanism have predicted
that the proportion
of fast and bright novae -- like Type Ia SNe, which are thought to have
similar progenitors -- is expected to be higher in younger stellar
populations that contain on average more massive white dwarfs (Yungelson
et al. 1997).
Since novae with massive
white dwarfs are expected to
have shorter recurrence times between eruptions,
the luminosity-specific nova rate (LSNR)
should also be higher in younger populations.
Despite these predictions, the weight of observational evidence
suggests the opposite; that the LSNR is at least as high (and perhaps higher)
in ellipticals and in spiral bulges, compared with early-type galaxies
(e.g., Williams \& Shafter 2004).

The most throughly studied extragalactic system is M31, where
more than 700 novae have been discovered since the pioneering work
of Hubble early in the 20th century (e.g. see Darnley et al. 2006; Pietsch
et al. 2007; Shafter 2007a, and references therein).
A principal conclusion reached
by these studies is that,
contrary to the predictions of population synthesis models,
M31 novae appear to belong primarily to the galaxy's bulge population.
This surprising result led
Ciardullo et al. (1987) to speculate that the bulge nova rate may
be enhanced by nova binaries that were spawned in M31's
globular cluster system,
and subsequently injected into the bulge through 3-body
encounters in clusters (McMillan 1986),
or by tidal disruption of entire clusters, or both.

To test the possibility that white dwarf accretors are also enhanced
in the clusters themselves, searches for novae
in M31's globular cluster system were undertaken
by Ciardullo et al. (1990a) and by Tomaney \& Shafter (1992).
In the former study, Ciardullo et al. examined 54 globular clusters
that fell within their prior M31 bulge nova surveys (Ciardullo et al.
1987; Ciardullo et al. 1990b). They found no nova eruptions, and
were able to show that novae were several times less likely to be found
in globular clusters than were high-luminosity X-ray sources,
where roughly $\sim$20\% are associated with globular clusters
(Crampton et al. 1984).
In the latter study, Tomaney \& Shafter used a multi-fiber spectrograph
to monitor more than half ($\grtsim$200) of M31 globular clusters over an
effective survey time
of almost a year. Once again, no novae were found suggesting that any
enhancement of novae in M31's globular clusters was less than that
of the high-luminosity X-ray sources.

The situation changed recently when as part of the ROTSE-IIIb program,
Quimby et al. (2007) discovered a nova, M31N-2007-06b, that was
spatially coincident with the M31 globular cluster system Bol 111.
In this Letter we report spectroscopic observations
of M31N-2007-06b, which clearly establish that the nova
belongs to the He/N spectroscopic class proposed by Williams (1992).
After describing the spectroscopic observations in the next section,
we conclude by discussing the
implications of an He/N nova arising in a globular cluster.

\section{Observations}

We discovered nova M31N-2007-06b on June 19.38 UT as part of the Texas
Supernova search (Quimby 2006; see Figure 1). The
nova was found at an unfiltered magnitude of $16.96 \pm 0.07$
(calibrated against the USNO-B1.0 R2) after removal of the static
background using a modified version of the PSF-matched image
subtraction code supplied by the Supernova Cosmology
Project (Perlmutter et al. 1999). It was $16.89 \pm 0.06$ mag on June
21.38, suggesting the discovery came near, and likely just prior to
maximum light. It was not present to a limiting magnitude of 17.88 on
June 15.38. By June 30.33 the nova had faded to below 17.89, placing a
lower limit on the fade rate of $>0.11$ mag per day.
M31N-2007-06b is located at $\alpha=00^h42^m33\fs13$,
$\delta=+41\arcdeg00\arcmin26\farcs3$ (J2000.0; $\pm 0.4''$ in each
coordinate). This position is consistent to within the errors of the
core of the cataloged M31 globular cluster
Bol~111 (Galleti et al. 2004), which appears to be relatively old
and metal poor (Jiang et al. 2003).

Spectroscopic observations were carried out with the Low Resolution
Spectrograph (LRS; Hill et al.1998) on the 9.2m Hobby-Eberly
Telescope (HET) beginning June 22.44. We used the $g1$ grating with a
2.0$''$ slit and the GG385 blocking filter, which covers
4150-11000\,\AA\ with a resolution of $R\sim 300$, although we limit
our analysis to the 4150-8900\,\AA\ range where the effects of order
overlap are minimal. Additional HET/LRS spectra were obtained on June
25.42 and July 25.34 UT with the $g2$ grism ($R\sim 650$ covering
4300-7300\,\AA).

Our initial spectrum, which was taken roughly three days after
eruption, is shown in Figure~2.  The spectrum is characterized by
strong and broad (FWMH$\simeq3000$~km~s$^{-1}$) Balmer emission lines,
as well as permitted lines of singly and doubly-ionized nitrogen at
5001\,\AA\ and 4640\,\AA, respectively. The spectrum is clearly that
of a classical nova. The presence of broad Balmer, NII and NIII
emission lines coupled with the absence of significant Fe~II emission
features establish that the nova is a member of the He/N class in
the system of Williams (1992). The significance of this classification
will be discussed in section~3.2 below.

The higher resolution follow-up spectra, which were obtained $\sim 1$
week and $\sim 5$ weeks post eruption, are shown in Figure~3.
As the nova faded, absorption features
(e.g. H$\beta$, Mgb\ $\lambda$5167,5173,5183, and NaD\ $\lambda$5890,5896)
from the underlying globular cluster became visible in the spectrum,
providing us the opportunity to
compare the radial velocities of these features
with those of the nova emission lines.
If M31N-2007-06b erupted in Bol~111,
the radial velocities should be consistent, whereas in the (highly)
unlikely event of a chance superposition of the nova on the cluster,
no such agreement would be required.
Despite the fact that accurate (absolute) radial
velocities are not possible to achieve at our spectral resolution,
it is reassuring that
the measured emission-line velocity in our final spectrum, $-320$~km~s$^{-1}$
(based on H$\alpha$ and H$\beta$), is consistent with a
mean velocity of $-300$~km~s$^{-1}$ derived from the
globular cluster absorption features.\footnote{We note that
the published radial velocity of Bol~111 is $-414\pm14$km~s$^{-1}$
(Galleti et al. 2006). Most of the $\sim$100~km~s$^{-1}$ discrepancy with
our measurement is likely due to a small zero-point error in our absolute
wavelength calibration of order 1~pixel at our resolution.}
We conclude that the weight of the available
evidence strongly supports the conclusion that M31N-2007-06b
is indeed associated with Bol~111.

\section{Discussion}

\subsection{Are Novae Enhanced in Globular Clusters?}

It has long been recognized that the number of X-ray sources per unit
mass is of order a hundred to a thousand times higher
in globular clusters compared with the rest of the Galaxy
(Clark 1975; Katz 1975).
A similar enhancement of X-ray sources has been seen in M31's globular
cluster population (Crampton et al. 1984; Di Stefano et al. 2002;
Trudolyubov \& Priedhorsky 2004).
The realization that these X-ray sources are the result of
dynamical interactions involving neutron stars
has led to the expectation that globular clusters should produce an even
greater number of close red dwarf -- white dwarf binaries, including
novae (Hertz \& Grindlay 1983; Rappaport \& Di Stefano 1994).
After initially disappointing searches, recent observations with
{\it HST\/} and {\it Chandra\/} have started to reveal increasing numbers
of these binaries in Galactic globular clusters (Edmonds et al. 2003;
Heinke et al. 2003; Knigge et al. 2002; Pooley et al. 2002).
To date at least one, and likely two classical novae have been observed
in the cores of Galactic globular clusters: T~Sco in M80 (Luther 1860;
Pogson 1860; Sawyer 1938), and an anonymous nova near the core of
M14 in 1938 (Hogg \& Wehlau 1964).
More recently, Shara et al. (2004) have discovered
a nova coincident with a globular cluster of M87.

Taken at face value, the
discovery of a single nova in one of M31's $\sim$300
globular clusters over the past $\sim$90 years yields
a rough lower limit on the frequency
of novae in globular clusters of, $f\sim4\times10^{-5}$ novae per cluster
per year. This result is consistent with a what one would expect if
novae were not enhanced with respect to the field, where
given that the M31 globular cluster system represents roughly 0.05\% of M31's
total mass in stars (Bramby \& Huchra 2001),
a global nova rate of $\sim50$ per year (Shafter \& Irby 2001;
Darnley et al. 2006) suggests that we
can expect to observe $\sim$2 novae per century
in M31's globular cluster population.
However, it is clear that our
lower limit on $f$ is unrealistically conservative.
In particular,
the effective survey time is significantly less than 90 years, as
M31 has been monitored only sporadically during this time.
In addition, novae are considerably more difficult to detect
against the background cluster light than they are in the field.
Thus, it is likely that many globular cluster novae have gone
undetected, making it likely
that the specific frequency of novae in globular
clusters is significantly higher than that in the field.

We conclude this discussion by noting that the number of nova
binaries in globular clusters may be significantly higher than
the observed frequency of nova eruptions would suggest.
In particular, the discussion above implicitly assumes that
the mean recurrence time for novae in globular clusters is the
same as that for field novae, whereas this may not be the case.
As shown by numerous studies, the frequency of nova events
depends primarily on three parameters: the mass accretion rate,
and the mass and temperature of the white dwarf (Livio 1992,
Townsley \& Bildsten 2005).
For a given mass accretion rate, nova systems with relatively massive
white dwarfs will have shorter recurrence times between eruptions
because less accreted mass is required to trigger a TNR.
Furthermore, as shown by
Townsley \& Bildsten (2005), the recurrence times are particularly
sensitive
to $T_{wd}$, with significantly less accreted envelope mass
require to achieve a TNR on hotter white dwarfs.
Since the accretion rate and white dwarf temperature are coupled,
the observed nova rates are essentially a function of the
mass transfer rate and white dwarf mass.
If these parameters differed systematically between novae in different
stellar populations, with an older population, for example, containing
on average cooler
and less massive white dwarfs, then the 
number of nova binaries
in globular clusters could be significantly higher than the
number of observed outbursts suggests.

\subsection{Nova Populations}

In recent years there has been a growing body of evidence that
there indeed exist two distinct populations of novae.
Duerbeck (1990) was first to formally postulate the existence of two
nova populations: a relatively young population that he called ``disc novae",
associated with novae
found in the solar neighborhood and in the LMC, and ``bulge novae",
which were concentrated towards the Galactic center and the bulge of M31, and
characterized by generally slower outburst development. Shortly
thereafter, Della
Valle et al. (1992) showed that the average scale height above the Galactic
plane for fast novae appeared smaller than for novae with slower rates of
decline, and
Williams (1992) noted that Galactic novae could be divided into two
classes based on their spectral properties:
specifically, the relative strengths of the Fe~II and He and N emission
lines. Novae with prominent Fe~II lines (the Fe~II novae) usually show
P~Cygni absorption profiles, tend to evolve more slowly, have lower
expansion velocities, and have a lower level of ionization
compared with novae that exhibit strong lines of He and N (the He/N novae).
In addition, these He/N novae often display very strong neon lines,
which suggests that the seat of the eruption is a relatively
massive ONe white dwarf. Such novae do not appear to produce the
copious carbon-rich dust that is often formed in nova ejecta arising
from the lower mass CO white dwarfs (Gehrz et al. 1998).
Additional support for the two-population scenario was
provided by Della Valle \& Livio (1998), who looked into the
spatial distribution of Galactic novae with known spectral
class and reliable distance estimates. From a sample of 27 novae
they noted that Galactic novae that could be
classified as He/N were more concentrated to the Galactic plane,
and tended to
be faster and more luminous compared with their Fe~II counterparts.

Following these studies, 
a general picture (summarized in Table~1) has begun to emerge
where the disk novae are thought to
arise on generally more massive (often ONe) white dwarfs associated
with systems found in younger stellar populations. Since less
accreted mass is required to achieve a TNR on a more massive
white dwarf, typical disk novae would then be expected to
expel less material and to evolve
more quickly than their bulge counterparts with lower mass white dwarfs
(Della Valle \& Livio 1998; Livio 1992).
Furthermore, the smaller amount of accreted matter required
to achieve a TNR, will lead to shorter recurrence times
for novae with massive white dwarfs (e.g., Gil-Pons et al. 2003).
The fact that Galactic recurrent novae appear to have massive
white dwarfs is consistent with this picture (Hachisu \& Kato 2001). 
Since the nova ejecta is thought to be contaminated with material
dredged up from the underlying white dwarf, it is generally
accepted that novae arising on
CO white dwarfs produce carbon-rich ejecta, and eventually carbon-rich dust,
while novae arising on the more massive ONe white dwarfs evolve to
become ``neon novae", characterized by significant Ne line emission
(Gehrz et al. 1985).
Finally, according to Williams (1992), systems that eject a relatively
small amount of mass in a thin shell tend to display the higher
excitation He/N spectra, while those that eject mass in a continuous
wind produce lower excitation Fe~II spectra with P~Cygni features.

Given the expectation that He/N novae are thought to be
primarily associated with younger stellar populations,
it may seem surprising that
the first nova identified in an M31 globular
cluster is a member of this class. However, as in the field,
selection effects strongly influence the discovery of novae
in globular clusters.
Further, since novae with massive white dwarfs are more luminous
and have shorter recurrence times
compared with low mass systems,
it should not
be surprising that a significant fraction of the
{\it observed} globular cluster novae harbor massive white dwarfs
(and show He/N spectra), even if such systems are less common
in globular clusters than they are in the field.
In M31 generally,
$\sim15$\% of M31 novae with measured spectra
fall into the He/N class (Shafter 2007b), while
available evidence suggests that the
percentage of He/N novae in the Galaxy is even higher, $\sim$20-25\%
Shafter (2007, in preparation). If this difference is found to be
statistically significant, it is possible that
the slightly higher fraction of He/N
novae observed in the Galaxy reflects the slightly later
Hubble type of the Milky Way compared with M31.

\section{Conclusions}

Our principal conclusions can be summarized as follows:

1) The nova M31N-2007-06b is spatially coincident to within
observational errors ($\pm 0.4''$ in each
coordinate) with
one of the galaxy's globular clusters (Bol 111).
This discovery marks the first time a nova has been associated
with a globular cluster in M31, and only the second extragalactic
nova to arise in a globular cluster
(the other being the nova found by Shara et al. in one of M87's
globular clusters.)

2) Our spectroscopic observations clearly establish that
M31N-2007-06b is a member of the He/N type in William's (1992) classification
system. Although such novae are expected to be more common in
younger stellar populations like that associated with M31's disk,
the shorter recurrence times and the higher luminosities expected for
these novae strongly enhance their discovery probability
in any stellar population.

3) Taken at face value, the association of a single nova with a
globular cluster out of the more than 700
novae discovered in M31 over the past century is consistent with the
hypothesis that novae in globular clusters are not enhanced
relative to the field.
However, it is almost certain that
a significant number of novae in M31's globular cluster
system have been missed over the past century
given the difficulty
in detecting globular cluster novae, making an enhancement over the field
seem likely. However, a more thorough and systematic monitoring campaign
of M31's globular cluster system over many years will be required
before a definitive conclusion can be reached.

\acknowledgments

We acknowledge support from NSF grants AST-0607682 (AWS) and AST-0707769 (RMQ).

{\it Facilities:} \facility{HET}.






\clearpage

\begin{deluxetable}{lcc}
\tablenum{1}
\tablewidth{0pt}
\tablecolumns{3}
\tablecaption{Putative Properties of Nova Populations}
\tablehead{& \colhead{Bulge Population} & \colhead {Disk Population}}
\startdata
Environment & Pop~II & Pop~I \\
Light Curve Evolution & Slow & Fast \\
Mean Recurrence Times & Longer & Shorter \\
White Dwarf Composition & CO & ONe \\
Dust Formation? & Yes & No \\
Williams (1992) Spectral Class & Fe~II & He/N \\
\enddata
\end{deluxetable}



\clearpage




\begin{figure}
\includegraphics[angle=0,scale=.55]{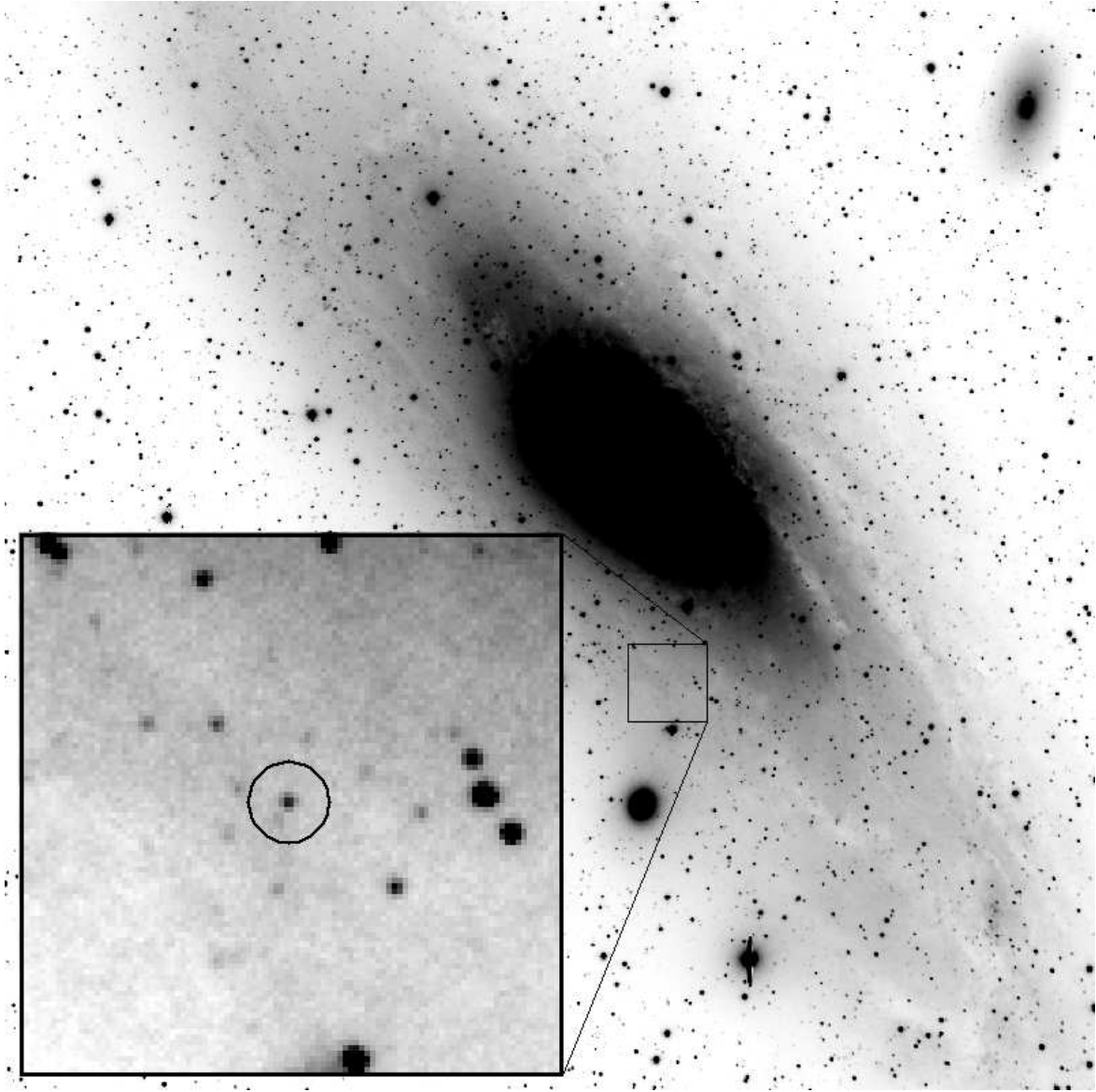}
\caption{Messier 31 as observed by ROTSE-IIIb. The large image is a
  co-addition of 103 images taken between 2004 Dec. and June 2006, and
  the inset shows the ROTSE-IIIb data from 2007 June 19. The light
  from the globular cluster is visible in the reference image (boxed).
\label{fig1}}
\end{figure}

\begin{figure}
\includegraphics[angle=270,scale=.65]{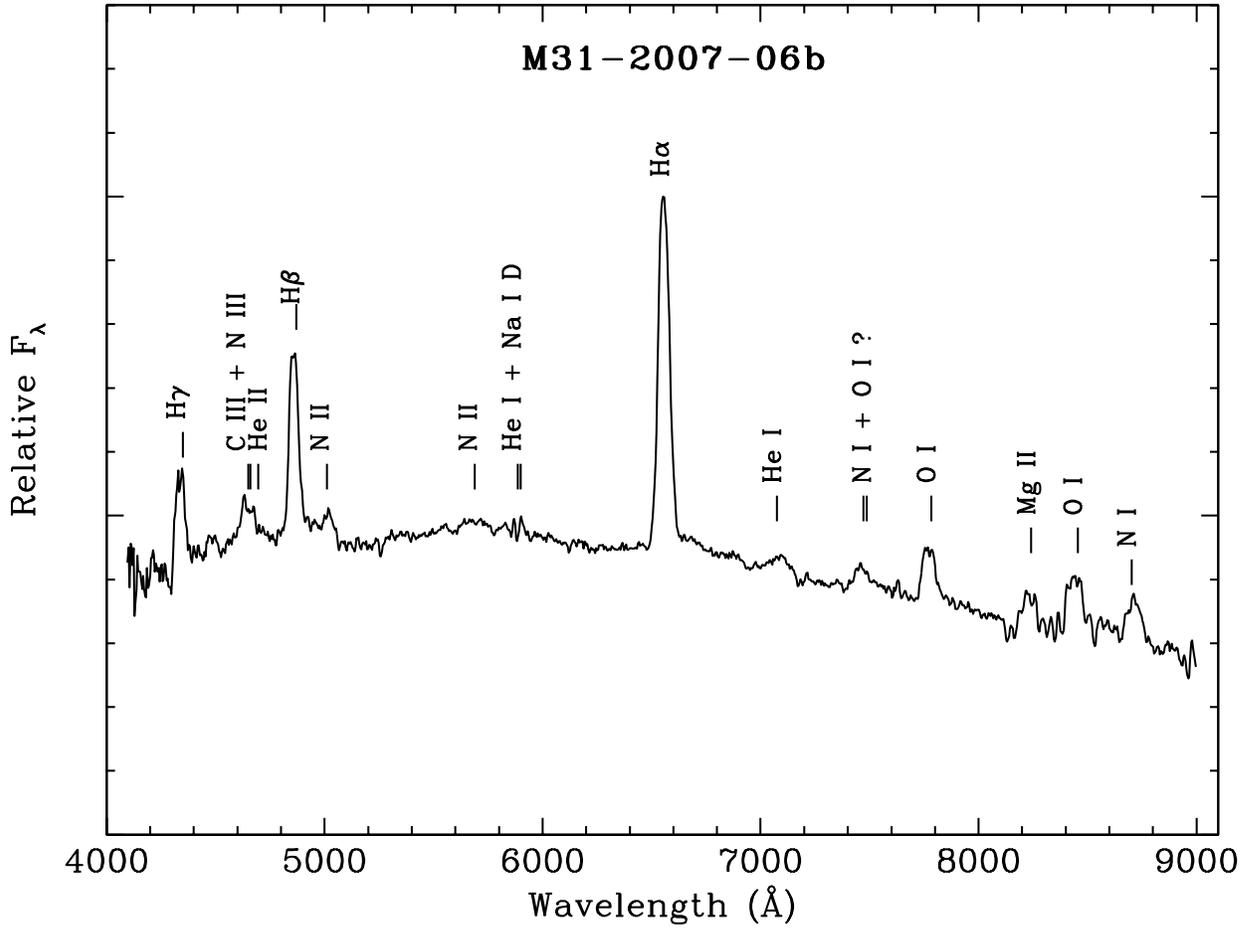}
\caption{A low-resolution spectrum of M31N2007-06b taken with the LRS on the
HET three days after discovery on 18~June~2007.
Note the broad Balmer, N~II and N~III lines, and the lack of
significant Fe~II emission characteristic of the He/N class of novae.
\label{fig2}}
\end{figure}

\begin{figure}
\includegraphics[angle=270,scale=.65]{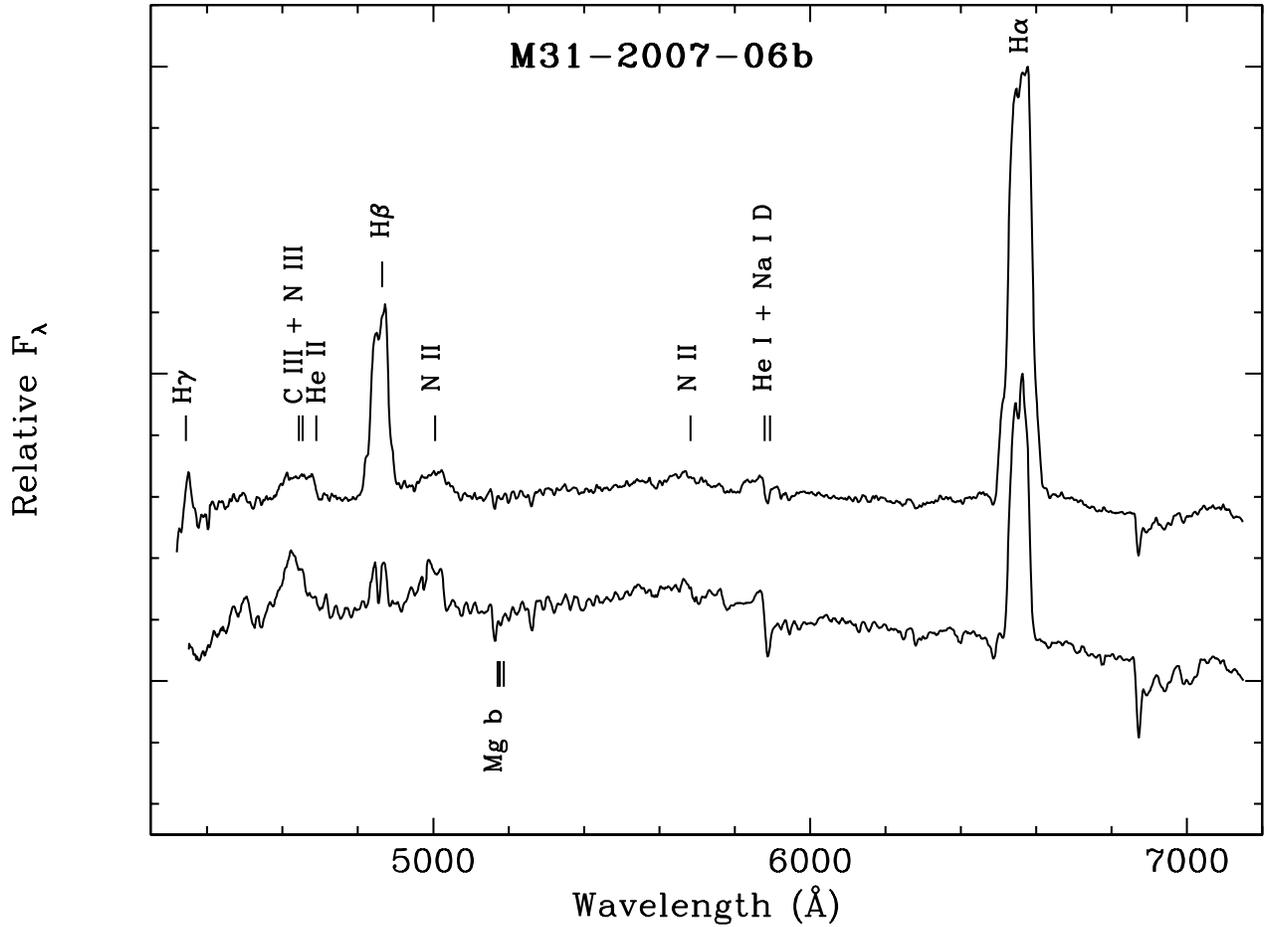}
\caption{Follow-up spectra of M31B-2007-06b taken with the LRS on the HET
(upper spectrum -- $\sim$1 week post eruption; lower spectrum -- $\sim$5 weeks
post eruption).
As the nova fades, note the increasing contamination by light from the
globular cluster, Bol 111, resulting in Balmer and Mg~b absorption
features.\label{fig3}}
\end{figure}





\end{document}